\documentclass[10pt]{article}
\usepackage[dvips]{graphicx}
\usepackage{wrapfig}
\begin{document}
\noindent{\small UDK 61.80.Az, 74.25.Fy, 74.72.Bk, 02.70.Bf}\\

\begin{center}
\Large{\bf NUMERICAL INVESTIGATION OF SCALING REGIMES IN A MODEL OF
ANISOTROPICALLY ADVECTED VECTOR FIELD}\\
\bigskip
\large{\it E. Jur\v{c}i\v{s}inov\'a,$^{1,2}$ M. Jur\v{c}i\v{s}in,$^{1,3}$ R. Remecky,$^{4}$ and M. Scholtz$^{4}$}\\
\smallskip
\small{$^{1}$ Institute of Experimental Physics, SAS, Watsonova 47, 04001 Ko\v{s}ice, Slovakia \\
$^{2}$ Laboratory of Information Technologies, JINR, 141 980 Dubna, Moscow Region, Russia \\
$^{3}$ Bogoliubov Laboratory of Theoretical Physics, JINR, 141 980
Dubna, Moscow Region,
Russia \\
$^{4}$ Department of Physics and Astrophysics, Institute of Physics,
P.J. \v{S}af\'arik University, \\ Park Angelinum 9, 04001
Ko\v{s}ice, Slovakia}
\end{center}

{\small{Influence of strong uniaxial small-scale anisotropy on the
stability of inertial-range scaling regimes in a model of a passive
transverse vector field advected by an incompressible turbulent flow
is investigated by means of the field theoretic renormalization
group. Turbulent fluctuations of the velocity field are taken in the
form of a Gaussian statistics with zero mean and defined noise with
finite correlations in time.  It is shown that stability of the
inertial-range scaling regimes in the three-dimensional case is not
destroyed by anisotropy but the corresponding stability of the
two-dimensional system can be corrupted by the presence of
anisotropy. A borderline dimension $d_c$ below which the stability
of the scaling regime is not present is calculated as a function of
anisotropy parameters.}}

\bigskip

\begin{center}
\bf INTRODUCTION
\end{center}

It is well known that the breakdown of the classical
Kolmogorov-Obuchov phenomenological theory of fully developed
turbulence \cite{Frisch} is more noticeable for simpler models of
passively advected scalar or vector quantity (scalar or vector
field) than for the velocity field itself. This phenomenon is
encoded in the terms intermittency and anomalous scaling
\cite{Frisch,MonYagBook}. At the same time, the problem of a passive
advection is easier from theoretical point of view (see, e.g.,
\cite{FaGaVe01} and references therein). Therefore it leads to the
fact that the problem of anomalous scaling can be understand here in
a few ways what is not possible for the present in the problem of
genuine turbulence.

One of the most effective approach for studying self-similar scaling
behavior is the method of the field theoretic renormalization group
(RG) \cite{ZinnJustin,Vasiliev}. It can be also used in the theory
of fully developed turbulence and related problems
\cite{Vasiliev,AdAnVa96,AdAnVa99}, e.g., in the problem of a passive
scalar (or vector) field advected  by a given stochastic
environment.

In \cite{AdAnVa98} the field theoretic RG was applied to the
so-called rapid-change model of a passive scalar advected by a
self-similar white-in-time velocity field which is also known as
Kraichnan model. It was shown that within the field theoretic RG
approach the anomalous scaling is related to the existence of
"dangerous" composite operators with negative critical dimensions in
the framework of the operator product expansion (OPE)
\cite{Vasiliev,AdAnVa96,AdAnVa99}.

Afterwards, various generalized descendants of the Kraichnan model,
namely, models with inclusion of large and small scale anisotropy,
%\cite{AdAnHnNo00},
compressibility,
%\cite{AdAn98}
and finite correlation time of the velocity field
%\cite{Antonov99,Antonov00}
were studied by the field theoretic approach (see \cite{Antonov06}
and references therein). Moreover, advection of a passive vector
field by the Gaussian self-similar velocity field (with and without
large and small scale anisotropy, pressure, compressibility, and
finite correlation time) has been also investigated and all possible
asymptotic scaling regimes and crossovers among them have been
classified \cite{all1,AdAnRu01,all2,AnHnHoJu03}. General conclusion
is: the anomalous scaling, which is the most important feature of
the Kraichnan rapid-change model, remains valid for all generalized
models.

In what follows we shall begin with investigation of one particular
model of a passive vector advected by a Gaussian velocity field with
finite correlation time in the presence of the small-scale
anisotropy, namely the model where the stretching term is absent
(the so-called $A=0$ model, see, e.g, \cite{AdAnRu01,AnHnHoJu03}).
This model is specific from several points of view but maybe the
most important fact is that in contrast to the other models of
passive vector admixture where the anomalous scaling is related to
the composite operators built of the vector field without
derivatives \cite{all2,AnHnHoJu03} in the case under consideration
it is related to the composite operators built solely of the
gradients of the field. This fact radically changes the complexity
of the problem especially in the anisotropic case (see, e.g.,
\cite{AdAnRu01,Novikov06} and references therein). Thus, in some
sense, it can be consider as a further step to the nonlinear
Navier-Stokes equation. In what follows, we shall present only the
first part of the RG analysis, namely, we shall analyze the
influence of the small-scale anisotropy on the infrared (IR)
stability of the possible scaling regimes of the model. It will be
seen that complexity of this task is also very close to the
corresponding problem in the stochastic Navier-Stokes equation
\cite{all3}.

\bigskip

\begin{center}
\bf 1. FIELD THEORETIC FORMULATION OF THE MODEL
\end{center}

We shall consider the model of the advection of transverse
(solenoidal) passive vector field ${\bf b} \equiv {\bf b}({\bf
x},t)$ which is described by the following stochastic equation
\begin{equation}
\partial_t {\bf b}  =  \nu_0 \Delta {\bf b} - ({\bf v \cdot \nabla}) {\bf b}   + {\bf f}, \label{K-K}
\end{equation}
where $\partial_t\equiv \partial/\partial t$, $\Delta \equiv{\bf
\nabla}^2$ is the Laplace operator, $\nu_0$ is the diffusivity (a
subscript $0$ denotes bare parameters of unrenormalized theory), and
${\bf v} \equiv {\bf v} ({\bf x} ,t)$ is incompressible advecting
velocity field. The vector field  ${\bf f} \equiv {\bf f} ({\bf x}
,t)$ is a transverse Gaussian random (stirring) force with zero mean
and covariance
\begin{equation}
D_{ij}^f \equiv \langle f_i({\bf x},t) f_j({\bf
x^{\prime}},t^{\prime}) \rangle= \delta(t-t^{\prime})C_{ij}({\bf
r}/L), \,\,\,\,\ {\bf r}={\bf x}-{\bf x^{\prime}} \label{cor-b}
\end{equation}
where parentheses $\langle...\rangle$ hereafter denote average over
corresponding statistical ensemble. The noise defined in
Eq.\,(\ref{cor-b}) maintains the steady-state of the system but the
concrete form of the correlator will not be essential in what
follows. The only condition which must be fulfilled by the function
$C_{ij}({\bf r}/L)$ is that it must decrease rapidly for $r\equiv
|{\bf r}| \gg L$, where $L$ denotes an integral scale related to the
stirring.

In real problems the velocity field ${\bf v}(x)$ satisfies
Navier-Stokes equation but, in what follows, we shall work with a
simplified model where we suppose that statistics of the velocity
field is given in the form of Gaussian distribution with zero mean
and pair correlation function
\begin{eqnarray}
&&\hspace{-1cm} \langle v_i(x) v_j(x^{\prime}) \rangle \equiv
D^v_{ij}(x; x^{\prime})= \int \frac{d^d {\bf k} d
\omega}{(2\pi)^{d+1}} R_{ij}({\bf k}) D^v(\omega,{\bf k})
e^{-i\omega(t-t^{\prime})+ i{\bf k}({\bf x}-{\bf x^{\prime}})},
\label{corv}
\end{eqnarray}
where $d$ is the dimension of the space, ${\bf k}$ is the wave
vector, and $R_{ij}({\bf k})$ is a transverse projector. In our
uniaxial anisotropic case it is taken as (see, e.g., \cite{all2} and
references therein)
\begin{equation}
R_{ij} ({\bf k})  =
%\langle v_i ({\bf x}, t) v_j (0,0) \rangle =
\left(1 + \alpha_{1} ({\bf n \cdot k})^2/k^2\right) P_{ij} ({\bf k})
+ \alpha_{2} n_s n_l P_{is} ({\bf k}) P_{jl} ({\bf k})\,,
\label{T-ij}
\end{equation}
where $P_{ij} ({\bf k})\equiv \delta_{ij}-k_i k_j/k^2$ is common
isotropic transverse projector, the unit vector ${\bf n}$ determines
the distinguished direction, and $\alpha_{1}$, $\alpha_{2}$ are
parameters characterizing the anisotropy. From the positiveness of
the correlation tensor $D^v_{ij}$ one immediately finds restrictions
on the values of the above parameters, namely $\alpha_{1,2}>-1$. The
function $D^v(\omega, {\bf k})$ in (\ref{corv}) is taken in the
following form \cite{AnHnHoJu03}
\begin{equation}
D^v(\omega, k) = \frac{g_0 u_0 \nu_0^3
k^{4-d-2\varepsilon-\eta}}{(i\omega+u_0 \nu_0
k^{2-\eta})(-i\omega+u_0 \nu_0 k^{2-\eta})}, \label{corrvelo}
\end{equation}
where $g_{0}$  plays the role of the coupling constant of the model
(a formal small parameter of the ordinary perturbation theory), the
parameter $u_{0}$ gives the ratio of turnover time of scalar field
and velocity correlation time, and the positive exponents
$\varepsilon$ and $\eta$ are small RG expansion parameters. The
coupling constant $g_{0}$ and the exponent $\varepsilon$ control the
behavior of the equal-time pair correlation function of velocity
field and the parameter $u_{0}$ together with the second exponent
$\eta$ are related to the frequency $\omega\simeq
u_{0}\nu_{0}k^{2-\eta}$ which characterizes the mode $k$. The value
$\varepsilon=4/3$ corresponds to the celebrated Kolmogorov
\char`\"{}two-thirds law\char`\"{} for the spatial statistics of
velocity field, and $\eta=4/3$ corresponds to the Kolmogorov
frequency. Simple dimensional analysis shows that $g_{0}$ and
$u_{0}$, which we commonly term as charges, are related to the
characteristic ultraviolet (UV) momentum scale $\Lambda$ (or inner
legth $l\sim\Lambda^{-1}$) by \begin{equation}
g_{0}\simeq\Lambda^{2\varepsilon},\qquad
u_{0}\simeq\Lambda^{\eta}.\end{equation}

The stochastic problem (\ref{K-K})-(\ref{corv}) can be treated as a
field theory with action functional \cite{ZinnJustin,Vasiliev}
\begin{eqnarray}
S(\Phi)&=&
%\int dt\,d^d{\bf x}\,\, \nonumber \\ &&
%\hspace{-0.7cm}
b_j^{\prime} \left[ \left(-\partial_t - v_i\partial_i+\nu_0\Delta +
\nu_0 \chi_{10} ({\bf n}\cdot{\bf
\partial})^2 \right)\delta_{jk} + n_j \, \nu_0 \left(\chi_{20} \Delta +
\chi_{30} ({\bf n}\cdot{\bf
\partial})^2 \right) n_k\right] b_k
\nonumber
\\ && - \frac{1}{2}
%\int dt_1\,d^d{\bf x_1}\,dt_2\,d^d{\bf x_2}
%\\ && \hspace{-0.7cm}
\left( v_i [D_{ij}^v]^{-1} v_j - b_i^{\prime} D^f_{ij} b_j^{\prime}
\right), \label{action2}
\end{eqnarray}
where $D_{ij}^v$ and $D^f_{ij}$ are given in (\ref{corv}) and
(\ref{cor-b}) respectively, ${\bf b}^{\prime}$ is an auxiliary
vector field (see, e.g., \cite{Vasiliev}), and the required
integrations over $x=({\bf x}, t)$ and summations over the vector
indices are implied. In action (\ref{action2}) the terms with new
parameters $\chi_{10},\chi_{20}$, and $\chi_{30}$ are related to the
presence of small-scale anisotropy and they are necessary to make
the model multiplicatively renormalizable. Model (\ref{action2})
corresponds to a standard Feynman diagrammatic technique (see, e.g.,
\cite{all2,all3} for details) and the standard analysis of canonical
dimensions then shows which one-irreducible Green functions can
possess UV superficial divergences.

The functional formulation (\ref{action2}) gives possibility to use
the field-theoretic methods, including the RG technique to solve the
problem. By means of the RG approach it is possible to extract
large-scale asymptotic behavior of the correlation functions after
an appropriate renormalization procedure which is needed to remove
UV-divergences.

\bigskip
\newpage

\begin{center}
\bf 2. SCALING REGIMES OF THE MODEL
\end{center}

\begin{wrapfigure}[23]{l}[0cm]{7.2cm}
\vspace{-1.5cm}
\includegraphics[width=6cm]{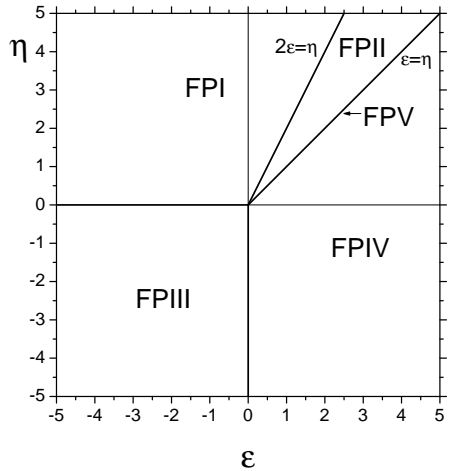}
\vspace{-0.8cm} \caption{\small The scaling regimes of the model in
the $\varepsilon-\eta$ plane. The region FPI corresponds to the
trivial rapid-change limit: $g^*/u^*=0, 1/u^*=0$. The region FPII
corresponds to the the non-trivial rapid change limit: $g^*/u^*>0,
1/u^*=0$. The region FPIII corresponds the trivial "frozen" limit:
$g^*=0, u^*=0$. The region FPIV is related to the non-trivial
"frozen" limit: $g^*>0, u^*=0$. In the end, the line
$\varepsilon=\eta$ (FPV) corresponds to the more interesting scaling
regime with $g^*>0, 0<u^*<\infty$.}\label{fig1}
\end{wrapfigure}

Details of the one-loop field-theoretic RG analysis of the model
will be done elsewhere. Here we only conclude that using the  RG
analysis leads to the following result: possible scaling regimes are
given by the IR stable fixed points of the system of five nonlinear
RG differential equations (flow equations or also known as
Gell-Mann-Low equations) for five scale dependent effective
variables (charges)
$\bar{C}=\{\bar{g},\bar{u},\bar{\chi}_1,\bar{\chi}_2,\bar{\chi}_3
\}$ of the model which are functions of the dimensionless scale
parameter $t=k/\Lambda$ \cite{Vasiliev,all3}. The system of the flow
equations is defined by the so-called $\beta$-functions  of the
model (they are functions of the charges, anisotropy parameters,
space dimension, and parameters $\varepsilon, \eta$) and it has the
following form
\begin{eqnarray}
t\frac{d {\bar g}}{d t}&=&\beta_g= \bar{g}
(-2\varepsilon+2\gamma_{1})\,,
\label{betagg}\\
t\frac{d {\bar u}}{d t}&=&\beta_u= \bar{u}(-\eta+\gamma_1)\,, \label{betauuu}\\
t\frac{d {\bar \chi_i}}{d t}&=&\beta_{\chi_i}= \bar{\chi}_i
(\gamma_1-\gamma_{i+1})\,, \label{betahh} \\
%\,\,\,\,\,
&& i=1,2,3\,,\nonumber
\end{eqnarray}
where the functions $\gamma_{i},i=1,2,3,4$ are given by the
following expressions (the one-loop approximation)
\begin{eqnarray}
\gamma_1&=&-g \frac{S_{d-1}}{(2\pi)^d}\frac{1}{2(d-1)(d+1)}
\int_{-1}^1 dx \frac{(1-x^2)^{(d-3)/2}}{w_1 w_2} K_1\,, \label{gammas}\\
\gamma_{i+1}&=&-\frac{g}{\chi_i}
\frac{S_{d-1}}{(2\pi)^d}\frac{1}{2(d-1)(d+1)} \int_{-1}^1 dx
\frac{(1-x^2)^{(d-3)/2}}{w_1 w_2} K_{i+1}\,,\quad i=1,2,3\,
\end{eqnarray}
where $S_d=2\pi^{d/2}/\Gamma(d/2)$ is the surface of the $d$
dimensional sphere, $w_1 = (1 + u + \chi_1 x^2), w_2 = (1 + u +
\chi_1 x^2 + (\chi_2 + \chi_3 x^2)(1 - x^2))$, and the coefficients
$K_{i}(x^2,g,u,\chi_1,\chi_2,\chi_3,\alpha_1,\alpha_2,d),i=1,2,3,4$
are some huge polynomials in respect to all variables and their
explicit form will be given elsewhere. The scale parameter $t$
belongs to the interval $0\leq t \leq 1$ with the initial conditions
given at $t=1$ and the IR stable fixed point corresponds to the
limit $t\rightarrow 0$, i.e., $\bar{C}|_{t=0}=C^*$.

\input epsf
   \begin{figure}[t]
     \vspace{-1.5cm}
       \begin{flushleft}
       \leavevmode
       \epsfxsize=6.2cm
       \epsffile{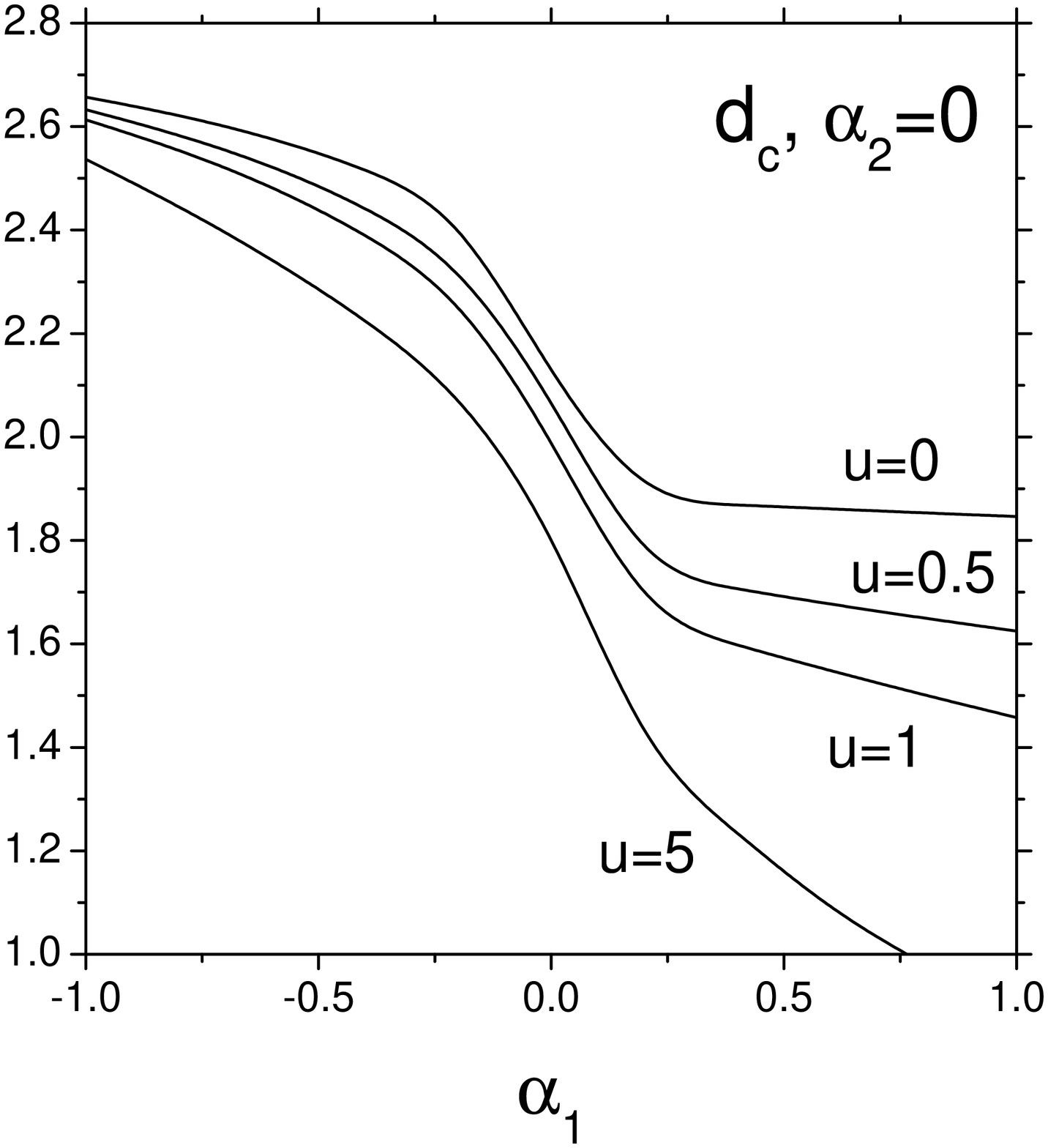}
   \end{flushleft}
     \vspace{-9.65cm}
   \begin{flushright}
       \leavevmode
       \epsfxsize=6.2cm
       \epsffile{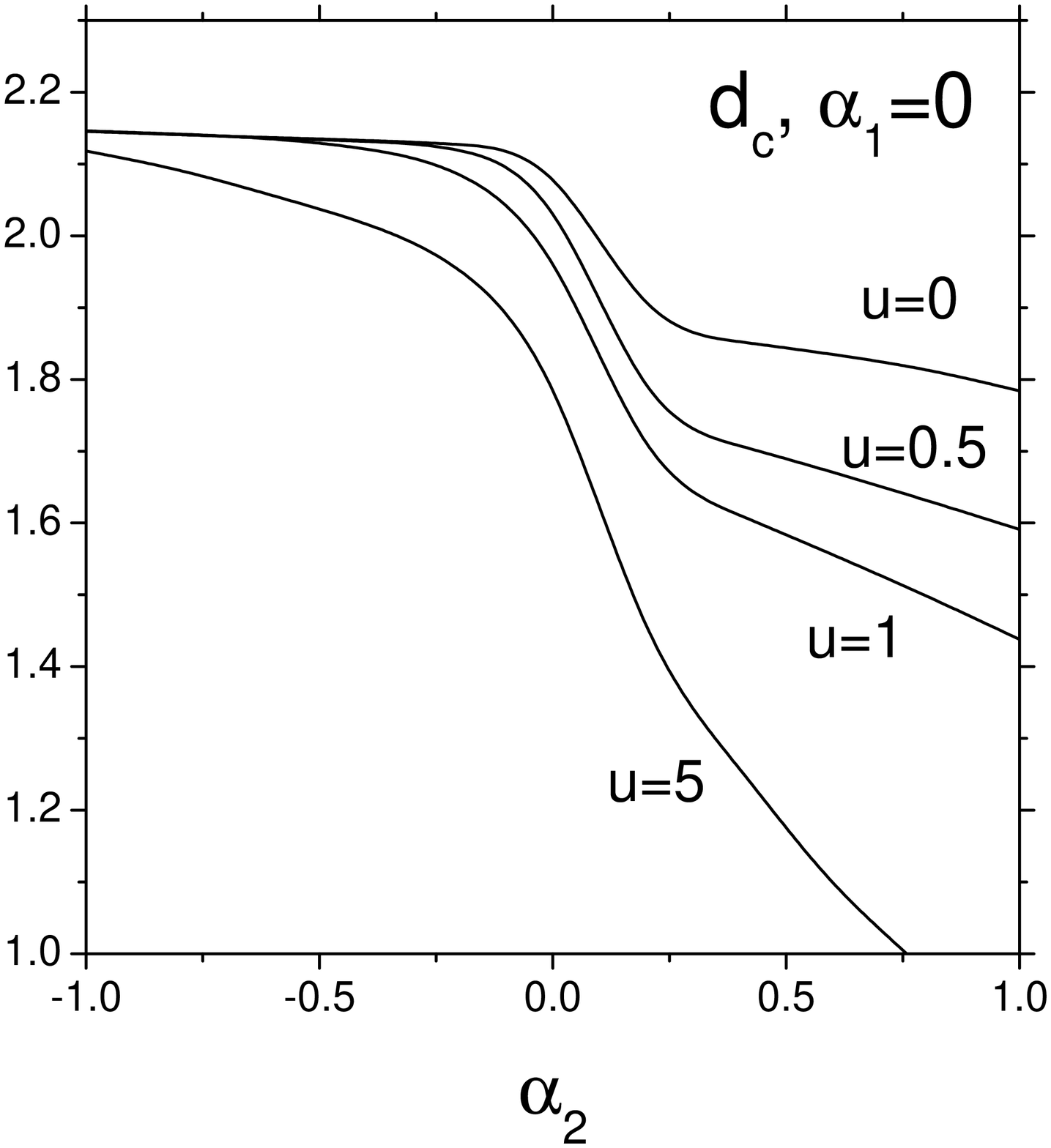}
   \end{flushright}
\vspace{-1.5cm} \caption{\small Dependence of the borderline
dimension $d_c$ on the parameters $\alpha_1$ and $\alpha_2$ for
different values of $u=u^*$. The corresponding scaling regime is
stable above the given curve. \label{fig2}}
\end{figure}

We have performed a numerical analysis of this system of
differential equations and our aim was twofold. First of all, we
have found all possible scaling regimes and we have analyzed the
regions of their IR stability in the $\varepsilon-\eta$ plane. The
results of this analysis are shown in Fig.\,\ref{fig1}, where it is
shown that the model exhibits five different scaling regimes (two
for rapid-change limit, two for so-called "frozen" limit, and one
general with nonzero $u_*$) (see, e.g., \cite{AnHnHoJu03} and
references therein). The second question which was investigated is
related to the dependence of stability of the above mentioned
scaling regimes on the anisotropy parameters $\alpha_1, \alpha_2$
and on the dimension of the space $d$. We have found the so-called
borderline dimension $d_c$ between stable and unstable regimes as a
function of anisotropy parameters $\alpha_1, \alpha_2$ and parameter
$u_*$. The results are shown in Fig.\,\ref{fig2} for some special
situations. One can see that the presence of small-scale anisotropy
leads to the violation of the stability of the corresponding scaling
regimes below $d_c\in[2,3]$ for appropriate values of anisotropy
parameters. But from the point of view of further investigation of
anomalous scaling the most important conclusion is that all the
three-dimensional scaling regimes remain stable under influence of
small-scale uniaxial anisotropy.

\bigskip

\begin{center}
\bf CONCLUSIONS
\end{center}

Using the field theoretic RG we have studied the influence of
small-scale uniaxial anisotropy on the stability of the scaling
regimes in the model of a passive vector advected by given
stochastic environment with finite time correlations. It is shown
that the system exhibits five different scaling regimes. They are
related to the values of the parameters $\varepsilon$ and $\eta$. On
the other hand, the stability of all these scaling regimes are
influenced by presence of small-scale anisotropy which is
demonstrated in the existence of the so-called borderline dimension
$d_c$ which is a function of the anisotropy parameters. The $d_c$ is
defined as dimension above which the corresponding scaling regime is
stable and below which the stability of the regime is destroyed.
All calculations have been done at the one-loop level. The results will be used
in the further investigations of the anomalous scaling of the model. \\

E.J. is thankful to J. Bu\v{s}a for discussion. The work was
supported in part by VEGA grant 6193 of Slovak Academy of Sciences
and by Science and Technology Assistance Agency under contract No.
APVT-51-027904.

%%%%%%%%%%%%%%%%%%%%%%%%%%%%%%%%%%%%%%%%%%%%%%%%%%%%%%%%%%%%%%

\def\refname{

\end{document}